\documentclass[
]{ceurart}
\usepackage{graphicx}

\usepackage{listings}
\usepackage{hyperref}
\hypersetup{colorlinks,allcolors=black}

\begin{document}

\copyrightyear{2024}
\copyrightclause{Copyright for this paper by its authors.
  Use permitted under Creative Commons License Attribution 4.0
  International (CC BY 4.0).}

\conference{PhysioCHI: Towards Best Practices for Integrating Physiological Signals in HCI,   May 11, 2024, Honolulu, HI, USA}

\title{CAN WE SAY A CAT IS A CAT? UNDERSTANDING THE CHALLENGES IN ANNOTATING PHYSIOLOGICAL SIGNAL-BASED EMOTION DATA}

\tnotemark[1]
%\tnotetext[1]{You can use this document as the template for preparing your
%  publication. We recommend using the latest version of the ceurart style.}

\author[1]{Pragya Singh}[%
email=pragyas@iiitd.ac.in,
url=https://alchemy18.github.io/pragyasingh/,
]
% \cormark[1]
\address[1]{IIIT-Delhi, Delhi, India}

\author[2]{Mohan Kumar}[%
email=mjkvcs@rit.edu,
url=https://www.rit.edu/directory/mjkvcs-mohan-kumar,
]
\fnmark[1]
\address[2]{Rochester Institute of Technology, Rochester, New York, US}

\author[3]{Pushpendra Singh}[%
email=psingh@iiitd.ac.in,
url=https://www.iiitd.ac.in/pushpendra/,
]
\fnmark[1]
\address[3]{IIIT-Delhi, Delhi, India}

% \cortext[1]{Corresponding author.}
\fntext[1]{Authors are advisors of this research work.}

\begin{abstract}
Artificial Intelligence (AI) algorithms, trained on emotion data extracted from physiological signals, provide a promising approach to monitoring emotions, affect, and mental well-being. However, the field encounters challenges because there is a lack of effective methods for collecting high-quality data in everyday settings that genuinely reflect changes in emotion or affect. This paper presents a position discussion on the current technique of annotating physiological signal-based emotion data. Our discourse underscores the importance of adopting a nuanced understanding of annotation processes, paving the way for a more insightful exploration of the intricate relationship between physiological signals and human emotions.
\end{abstract}


\maketitle
\section{Introduction}
Physiological signals have significant potential for widespread emotion monitoring, offering numerous applications in emotion-affect-mental-wellbeing recognition. The increasing prevalence of wearable devices equipped with sensors capable of continuously collecting physiological signal data has prompted researchers in the human-computer interaction (HCI) and artificial intelligence (AI) domains to explore their applicability in emotion-affect-mental-wellbeing monitoring and emotion-based interfaces. In contrast to methods relying on video, audio, or text for emotion recognition, leveraging physiological data provides several advantages. It enables unobtrusive data collection from individuals and groups over extended periods, offering high temporal resolution. Furthermore, the use of physiological data reduces susceptibility to conscious manipulation by subjects.

However, a well-labelled dataset is a crucial requirement for training AI algorithms in a supervised fashion. From past experience, it is evident that gathering physiological signals with their annotated emotional response is challenging. Prior work on the collection of emotion-affect-mental well-being datasets has explored both lab-based settings and field settings with constraint and daily-life activities \cite{schmidt2018introducing, koelstra2011deap, koldijk2014swell, hosseini2022multimodal}. Data collection for physiological signal-based data remains challenging, limiting the available datasets' capabilities for training AI algorithms \cite{9779458}. The major reason for these challenges lies in the nature of data. Physiological signal-based data is often collected by attaching sensors to the human body. However, the availability of portable wearable devices equipped with sensors is often limited. Another major challenge is the wearable devices equipped with the sensors often don't provide raw data, thus making it challenging to collect physiological data. Moreover, the nature of annotations in emotion-affect-mental well-being recognition is often not as straightforward as other application domains in AI; for example, in the computer vision dataset, we can label a picture of a cat as a cat with surety. The annotation procedure in this domain is heavily dependent on psychological models, emotion theories and human annotators, making the procedure of annotation complex and biased on various levels.
In this work, we present a preliminary discussion on the challenges in collecting and utilising annotations with respect to physiological signals for training emotion-affect-mental well-being recognition algorithms. Furthermore, we shed light on the adverse effect that present annotations have on overall data quality and model performance. Next, we present some avenues for HCI and AI researchers to overcome these challenges.

\section{Related Work}
Emotion recognition, a rapidly growing field within artificial intelligence and human-computer interaction, focuses on identifying and interpreting human emotions through various data modalities such as facial expressions, voice tone, physiological signals, and behavioural patterns. With the availability of wearable sensors, there is an increase in ongoing work in physiological signal-based emotion recognition \cite{10.1145/3361562, siirtola2023predicting, 10.1145/3460418.3479335}. Researchers have leveraged various physiological signals and their derived features, such as heart rate (HR), heart rate variability (HRV) measured using electrocardiogram (ECG), photoplethysmography (PPG) or blood volume pulse (BVP) signals, skin conductance measured using electrodermal activity (EDA), also known as galvanic skin response or resistance (GSR), skin temperature (SKT), and muscle activity measured using electromyogram (EMG) for quantifying changes in emotional arousal \cite{koelstra2011deap, schmidt2018introducing, koldijk2014swell, hosseini2022multimodal, abadi2015decaf}.
Furthermore, in the prior literature, physiological signal-based emotion datasets have been curated primarily in three settings: i) Laboratory settings \cite{schmidt2018introducing, subramanian2016ascertain}, ii) field with constraint settings \cite{hosseini2022multimodal, koldijk2014swell} and iii) real-life settings \cite{shui2021dataset, bota2024real}. For annotating the physiological data, prior works have employed techniques such as annotations by external annotators or experts \cite{healey_detecting_2005}, self-reports using self-assessment questionnaires such as Self-Assessment Manikin (SAM) or Positive and Negative Affect Schedule (PANAS) \cite{schmidt2018introducing}, experience sampling \cite{shui2021dataset}, continuous annotations by participants using annotation device \cite{sharma_dataset_2019, xue_rcea-360vr_2021, 8105870} and self-reflection \cite{10.1145/3334480.3383019}. Few other works have used the stimulation task as ground truth; for example, in the WESAD dataset \cite{schmidt2018introducing}, physiological data collected during the Trier Social Stress Test (TSST) is labelled as stress. Recently, there has been an increasing focus on collecting physiological data in everyday settings in order to capture realistic responses \cite{9779458}. While we see enough work in improving the setting of emotion data collection, limited attention is given to the quality of annotations and their effects on overall data quality and performance of emotion recognition algorithms. This position paper presents an argument on the limitations of present practices of annotating physiological signal-based emotion data.

\begin{figure}
  \centering
  \includegraphics[width=0.8\textwidth]{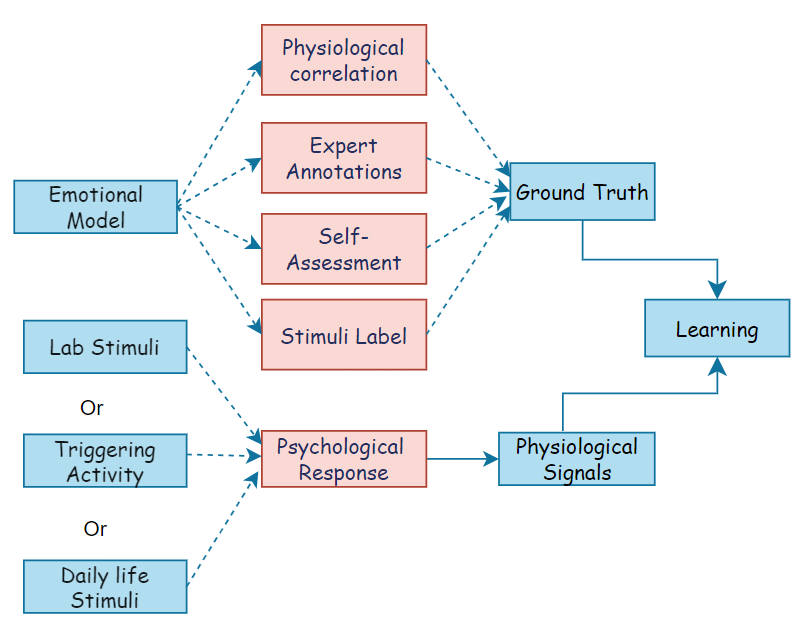}
  \caption{Presently used procedure of Physiological Signal based Emotion Data collection.}
  \label{fig:example}
\end{figure}

\section{Understanding the limitations in Annotation Procedure}
In this section, we have outlined the limitations present methods of annotating physiological signal-based emotion data suffer from and possible avenues to venture. 

\subsection{Annotating the bare-minimum}
Over the past decade, there have been multiple endeavours to recognize emotion-affect-mental well-being using various physiological signal data and AI. Although AI algorithms exhibit significant performance in laboratory settings with small sample sizes, identifying emotions, affect, or mental well-being in everyday settings remains challenging. Prior works on curating physiological signal-based emotion have primarily adopted three approaches for annotating physiological data. These approaches include: i) \textit{Self-assessment}: wherein emotions are annotated by the person experiencing the emotions, ii) \textit{Expert-annotations}: wherein emotions are annotated by an expert observer iii) \textit{Physiological Correlation}: wherein emotions are annotated based on the patterns or changes in physiological signals themselves. For example, specific patterns in heart rate variability and skin conductance might be associated with particular emotional states iv) \textit{Stimuli Label}: labels are assigned based on external stimuli or events used to elicit specific emotional responses. In figure \ref{fig:example}, we have illustrated the presently used methodology for physiological signal-based emotion data collection.

It is important to highlight that each of the above-mentioned methods suffers from certain limitations. For instance, self-reported annotations are often perceived emotions that can be based on underlying mood irrespective of the physiological changes and thus can lead to biased labels (For example, a person with an underlying sad mood may perceive an otherwise happy situation as sad due to his overall mood at the instance). Similarly, expert-based annotations are difficult in the case of real-life or ambulatory data collection. They may also suffer from expert bias (For example, an expert may annotate the emotions based on the manipulated expressions of a person while a person is not feeling them underneath). Further physiological correlations are often misleading in the case of ambulatory settings. They may also suffer from artefacts and collection environments (For example, electrodermal activity data may show peaks in case of a change in temperature). Subsequently, labels based on stimuli frequently fall short in capturing the subjective perception of diverse participants. These limitations underscore the necessity of devising a more nuanced approach to annotation that can delve into a comprehensive understanding of the larger context.

\subsection{Are Emotion Scales Enough?}
Another important aspect of emotion annotation is the choice of rating scale. Prior work has utilised scales inspired by either Emotional models or psychological tests for emotion-affect-mental well-being annotations. In prior literature, scales such as Self-Assessment-Manikins \cite{bynion2020self}, Geneva emotion wheel \cite{sacharin2012geneva}, PANAS \cite{watson1988development}, and others have drawn inspiration from emotion theories. Similarly, standard scales from psychology and psychiatry such as the Perceived Stress Scale (PSS) \cite{cohen1994perceived}, State-Trait Anxiety Inventory (STAI) \cite{marteau1992development}, General Health Questionnaire (GHQ) \cite{mccabe1996measuring}, have been utilized in the past for annotating physiological data. 
While employing standard questionnaires for annotating emotions proves convenient in controlled laboratory environments, it's essential to recognize the challenge of replicating such methods in everyday data collection settings, where soliciting responses from human participants can be challenging. Another critical consideration pertains to the design of current scales. Many of the prevalent scales are crafted to capture annotations over specific durations. Consequently, these scales are tailored for discrete annotations, such as those made after observing a stimulus, at the day's conclusion, or on a weekly basis, making them unsuitable for obtaining high-resolution continuous annotations. For example, a scale rooted in an emotional model like SAM excels at capturing overall valence and arousal within a given duration. However, in cases where there are numerous changes within the last hour, the annotation we obtain reflects the overall perceived emotion rather than providing a detailed, high-resolution account of those fluctuations. Another common observation is that the scales often suffer from cultural influence and definitions of human subjects' emotions in a region. Our observations from past work and our experiences suggest a need for designing shorter, continuous and self-descriptive scales for better annotations.  

\subsection{Can we say a Cat is a Cat?}
Prior work has highlighted that emotions are subjective, and so are physiological signals. Each individual has their own physiology and data range. The physiology of a person is also altered based on several bodily and environmental changes. The observed variations on multiple levels indicate the complexity of establishing a direct one-to-one mapping between physiological changes and emotions, affect, or mental well-being. Unlike in other domains leveraging AI, the annotation process in this context is not straightforward. This highlights the imperative to move beyond simple annotations and underscores the importance of not exclusively interpreting physiological changes as responses solely to alterations in emotions, affect, and mental well-being.

% \section{Proposed Solution}
% AI algorthimns powered by physiological signal based data present an avenue for continuous monitoring of emotions, affective states and mental well-being. As discussed above present methods of data collection suffer from several challenges limiting the progress in the field. In this section we propose few solutions to overcome these challenges:\\

% \textit{1) Who is your participant?}: In data collection procedures often participants are treated as belonging to similar background based on inclusion and exclusion criteria. While participants with diagnosed mental health conditions are mostly excluded until that study is collecting data for diagnosed patients, but the not diagnosed cohort is often treated as healthy. However with daily stress and hectic work and life environments that most people are living in globally may create different physiological thresholds even for healthy cohort. Thus we suggest a more nuanced approach to data collection wherein participant' psychological background, minor health history and their living conditions should also be included as a context.\\

% \textit{2) Designing low resolution scales}: Present scales are designed to capture emotional changes or affect changes within a duration, while the fluctuating nature is often not captured. We suggest a designing short descriptive scales to perceived emotions and short questionnaire to capture events in daily life that can lead to the perceived change if any.

\section{Conclusion}
In our paper, we presented a position discussion on the limitations present techniques of annotating emotion-affect-menatl well-being data suffer from. In the future, we want to explore these challenges further and develop solutions to overcome these limitations. In conclusion, comprehending the challenges associated with annotating physiological signal-based emotion data is imperative for advancing research in emotion recognition and understanding human affect. The complexities involved in accurately capturing and interpreting physiological responses underscore the need for innovative methodologies and robust annotation strategies. Addressing these challenges will enhance the reliability of physiological data and contribute to the development of more effective applications and interventions in fields such as healthcare, human-computer interaction, and psychology. 

\bibliography{sample-ceur}
\end{document}